# A quasi-RBF technique for numerical discretization of PDE's

## W. Chen
Department of Informatics, University of Oslo, P.O.Box 1080, Blindern, 0316 Oslo, Norway
e-mail: wenc@ifi.uio.no

Despite such very appealing features of the radial basis function (RBF) as inherent meshfree and independent of dimension and geometry, the various RBF-based schemes [1] of solving partial differential equations (PDE's) still confront some deficiencies. For instances, the lack of easy-to-use spectral convergent RBFs, ill-conditioning and costly evaluation of full interpolation matrix. The purpose of this study is a combined use of radial basis function and the other approximation methods to cure these drawbacks.

Atkinson [2] developed a strategy which splits solution of a PDE system into homogeneous and particular solutions, where the former have to satisfy the boundary and governing equation, while the latter only need to satisfy the governing equation without concerning geometry. Since the particular solution can be solved irrespective of boundary shape, we can use a readily available fast Fourier or orthogonal polynomial technique $O(N\log N)$ to evaluate it in a regular box or sphere surrounding physical domain.

The distinction of this study is that we approximate homogeneous solution with nonsingular general solution RBF as in the boundary knot method [3]. The collocation method using general solution RBF has very high accuracy and spectral convergent speed and is a simple, truly meshfree approach for any complicated geometry. More importantly, the use of nonsingular general solution avoids the controversial artificial boundary in the method of fundamental solution [4] due to the singularity of fundamental solution.

Due to the fact that the present scheme is a combination of the RBF and other approximation techniques, we call it the quasi-RBF method (QRM) compared with a pure RBF method. Experimenting the QRM with typical Laplace, Helmholtz and convection-diffusion problems shows that the method produces very accurate solutions with a relatively small number of nodes. It is worth stressing that the QRM converges very fast and is quite easy to learn and program, especially for high-dimensional complicated-shape problems.